\documentclass[submission,copyright,creativecommons]{eptcs}
\providecommand{\event}{QPL 2015} 
\usepackage{breakurl}             
\usepackage{graphicx}
\usepackage{amssymb}
\usepackage{amsmath, cancel}
\usepackage{mathtools}
\usepackage{epstopdf}
\usepackage{mathrsfs}

\title{Invariant Set Theory: Violating Measurement Independence without Fine Tuning, Conspiracy, Constraints on Free Will or Retrocausality}
\author{T.N.Palmer
\institute{Department of Physics\\ University of Oxford, United Kingdom}
\email{tim.palmer@physics.ox.ac.uk}
}
\def\titlerunning{Invariant Set Theory}
\def\authorrunning{T.N.Palmer}
\begin{document}

\maketitle

\begin{abstract}
Invariant Set (IS) theory is a locally causal ontic theory of physics based on the Cosmological Invariant Set postulate that the universe $U$ can be considered a deterministic dynamical system evolving \emph{precisely} on a (suitably constructed) fractal dynamically invariant set in $U$'s state space. IS theory violates the Bell inequalities by violating Measurement Independence. Despite this, IS theory is not fine tuned, is not conspiratorial, does not constrain experimenter free will and does not invoke retrocausality. The reasons behind these claims are discussed in this paper. These arise from properties not found in conventional ontic models: the invariant set has zero measure in its Euclidean embedding space, has Cantor Set structure homeomorphic to the p-adic integers ($p \ggg 0$) and is non-computable. In particular, it is shown that the p-adic metric encapulates the physics of the Cosmological Invariant Set postulate, and provides the technical means to demonstrate no fine tuning or conspiracy. Quantum theory can be viewed as the singular limit of IS theory when when $p$ is set equal to  infinity. Since it is based around a top-down constraint from cosmology, IS theory suggests that gravitational and quantum physics will be unified by a  gravitational theory of the quantum, rather than a quantum theory of gravity. Some implications arising from such a perspective are discussed. 

\end{abstract}

\section{Introduction}

Invariant Set (IS) theory \cite{Palmer:2014} \cite{Palmer:2015a} \cite{Palmer:2015b} is a putative deterministic locally causal theory of physics based on the Cosmological Invariant Set Postulate: the universe $U$ can be considered a locally causal deterministic dynamical system evolving \emph{precisely} on a fractal dynamically invariant set $I_U$ in $U$'s state space. Fractal invariant sets are features of a generic class of chaotic dynamical systems (e.g. \cite{Strogatz}). Like general relativity, IS theory proposes that the laws of physics are geometric - describing the geometry of state space in particular. Less like general relativity, the fractal geometry of IS theory links directly to aspects of number theory, exploited below. IS theory implies a much greater synergy between the physics of the very large and the very small than exists in contemporary physical theory. As such, IS theory has significant implications for the way we can understand the essential phenomena of quantum physics. The implications for the interpretation of the Bell Inequalities, and hence quantum nonlocality, are the focus of this paper. In particular, we attempt to show that IS theory is consistent with the a realistic locally causal explanation of quantum physics. 

A key parameter $N \gg 1$ in IS theory describes the fractal dimension of $I_U$. The larger is $N$, the closer this dimension approaches an integer (i.e. the closer the geometry is Euclidean). As discussed in \cite{Palmer:2015b}, for any finite $N$ there exists a 1-1 injection between the (symbolically defined) state space of IS theory and elements of the complex Hilbert space of quantum theory. The fact that this mapping is not a bijection means that certain properties of quantum theory, notably those associated with the algebraically closed nature of the complex Hilbert Space, do not hold in IS theory for any finite $N$. As a result, quantum theory emerges as the singular (and not the smooth) limit of IS theory at $N= \infty$. Singular limits are commonplace in physical theory \cite{Berry}: the inviscid Euler equations are the singular limit of the viscous Navier-Stokes equations at infinite Reynolds number, and classical physics is the singular limit of quantum theory when Planck's constant is set to zero. This has profound implications, discussed below. 

In Section \ref{MI}, we discuss how IS theory violates Measurement Independence. Then, in Sections \ref{ft}, \ref{conspire}, \ref{free} and \ref{retro}, we discuss, respectively, why such violation is not fine tuned, is not conspiratorial, is consistent with experimenter free will and does not invoke retrocausality.  In Section \ref{cosmo} we discuss the links between IS theory, cosmology, quantum measurement and the goal of unifying gravitational and quantum physics. Concluding remarks are made in Section \ref{conc}

\section{Violating Measurement Independence in Invariant Set Theory}
\label{MI}

As mentioned, IS theory can be considered a realistic locally causal ontic theory. Such a theory can only violate Bell's inequalities by violating (at least partially) the measurement independence condition \cite{Hall:2010} \cite{Hall:2011} \cite{Gallicchio:2014}
\begin{equation}
\rho(\lambda|\mathbf{a}, \mathbf{b}) = \rho(\lambda| \mathbf{c}, \mathbf{d})
\end{equation}
for all measuring orientations $\mathbf{a}, \mathbf{b}, \mathbf{c}, \mathbf{d}$, and where the ontic variable $\lambda$, describes some (quantum) system subject to measurement. In general, physicists do not consider that so-called `superdeterministic' theories which violate measurement independence are physically plausible. Amongst the various objections raised are the implications of a very finely tuned theory \cite{WoodSpekkens}, of an implausible conspiracy between the values of the ontic variables and the determinants of instrument settings \cite{Bell} or of unacceptable constraints on experimenter free will. The purpose of this paper is to show that none of these objections apply to IS theory.

In IS theory, the measurement independence condition is violated in the following way:
\begin{equation}
\label{MD}
\rho(\lambda|\mathbf{a}, \mathbf{b}') \ne \rho(\lambda| \mathbf{a}, \mathbf{b})
\end{equation}
where $\mathbf{a}, \mathbf{b}, \mathbf{b'}$ are three points on $\mathbb{S}^2$ such that
\begin{equation}
\label{constraint1}
\cos \theta_{ab'} \in \mathbb{Q}_2(N), \;\; \cos \theta_{ab} \notin \mathbb{Q}_2(N)
\end{equation}
where $\cos \theta_{ab}$, denotes the angular distance between $\mathbf{a}$ and $\mathbf{b}$ etc and  $\mathbb{Q}_2(N)$ denotes the set of rational numbers describable by $N$ bits. In particular, IS theory draws on probability distributions $\rho$ where the left hand side of (\ref{MD}) is non-zero (corresponding to attributes of states $X_U$ of $U$ lying on $I_U$), whilst the right hand side is strictly zero (corresponding to attributes of certain counterfactual states of $U$ not lying on $I_U$). In addition to (\ref{constraint1}), IS theory also requires phase angles $\phi$, expressed as a multiple of $\pi$, to be describable by $N$ bits:
\begin{equation}
\label{constraint2}
\phi/\pi \in \mathbb{Q}_2(N)
\end{equation}
In seeking an interpreting the violation of measurement independence in IS theory, one should not imagine that measurement settings causally affect the ontic state of the system being measured, or vice versa. Rather, the fractal geometry $I_U$, whose description is presumed to define the laws of physics at their most primitive level, provides an overarching `top-down' constraint on possible joint distributions of measurement settings and ontic states.

The relevance of (\ref{MD}) for the (original) Bell inequalities is as follows. Let $\mathbf{a}$, $\mathbf{b}$ and $\mathbf{c}$ denote three distinct points on the sphere (corresponding to three orientations in physical space). The cosine rule applied to the spherical triangle $\triangle_{abc}$ gives
\begin{equation}
\label{triangle}
\cos \theta_{ab}=\cos \theta_{ac} \cos \theta_{bc}+\sin \theta_{ac} \sin \theta_{bc} \cos \phi
\end{equation}
where $\phi$ denotes the interior angle of $\triangle_{abc}$ at the apex $c$. The three cosines in (\ref{triangle}) all appear in Bell's inequality
$$
\label{bell}
|\text{Corr}_{\rho}(\mathbf{a}, \mathbf{b})-\text{Corr}_{\rho}(\mathbf{a}, \mathbf{c})| \le 1+\text{Corr}_{\rho}(\mathbf{b}, \mathbf{c})
$$
where $\text{Corr}_{\rho}(\mathbf{a}, \mathbf{b})=\cos \theta_{ab}$ etc. As mentioned, in IS theory not only do we require that each of these cosines is describable by $N$ bits, but so also is the angle $\phi$, expressed as a multiple of $\pi$ (the reader is referred to \cite{Palmer:2015b} for a discussion of the fractal geometric reason behind these constraints). It is easy to show, by rudimentary number theoretic arguments, that this condition is simply not compatible with the cosine rule above. For example, with $0<\phi<\pi/2$, then if $\cos \theta_{ac}$ and $\cos \theta_{bc}$ are rational, so that the first term on the right hand side of (\ref{triangle})) is rational, then the second term on the right hand side of (\ref{triangle}) will be irrational and so the left hand side as a whole will be irrational. Hence IS theory is not itself constrained by the Bell inequality. 

As discussed in more detail in Section \ref{ft} in the context of the CHSH version of the Bell inequality, IS theory predicts that experimenters never actually test (\ref{bell}) - or the corresponding CHSH inequality - but rather test a modified version of this inequality where all the individual correlations are describable by $N$ bits. Such a modified version is experimentally testable (i.e. corresponds to states $X_U$ on the invariant set) because the three individual correlations can then be foud by three separate sub-experiments, each on the invariant set, and each with measurement orientations having rational cosines. 

Making a crucial distinction between rationals and irrationals in this way is likely to bring a sense of profound uneasiness to a reader groomed in Euclidean (more generally Riemannian and pseudo-Riemannian) geometry and real (or indeed complex) analysis: from these perspectives, it would seem that IS theory requires a level of precision that is completely unphysical. Using the formalism of p-adic numbers, relevant in describing fractal geometries, we attempt to show in the next Section that this interpretation is false. 

\section{Fine Tuning and the p-adic Numbers}
\label{ft}

Superficially, the violation of measurement independence as described by (\ref{MD}) appears to embody the undesirable concept of fine tuning \cite{WoodSpekkens}. Take a number  $\in  \mathbb{Q}_2(N)$ for $N \gg1$ and perturb it with an arbitrarily small number drawn randomly from the reals. Then almost certainly the perturbed number $\notin \mathbb{Q}_2(N)$. That is to say, IS theory's crucial distinction as to whether the cosine of an angle is or is not describable by $N \gg 0$ bits seems to be completely destroyed by adding the smallest amount of noise. 

A more careful analysis of this objection raises profound questions about the meaning of words such as `fine' and `small' in IS theory: as discussed in the Introduction, our intuition about these words is largely based on 19th Century notions of continuity jn Euclidean geometry. However, IS theory is based on the central importance of a fractal invariant set $I_U$ in the state space of the universe. To discuss how to formulate the notion of continuity (and indeed differentiability) in this context, note first that we can write, $I_U=\mathbb{R} \times C(p)$, where $C (p)$ 
$$
\label{CN}
C(p)=\bigcap_{k \in \mathbb{N}} C_k(p) \nonumber
$$
is a Cantor Set. For simplicity, consider $C(p)$ as a subset of the unit interval $[0,1]$. Specifically, let us suppose that the $k+1$th iterate $C_{k+1}(p)$ is generated from the $k$th iterate $C_k(p)$ by dividing an interval of $C_k(p)$ into $2p-1$ subintervals and removing every second subinterval. With $p=2$, $C(2)$ is the familiar Cantor Ternary Set. However, IS theory requires $p=2^N+1$ (with $N$ itself a power of 2). If there are only finitely many Fermat primes (e.g. as suggested in  \cite{HardyWright}), then it is tempting to speculate that the value of $N$ is set by the condition that $p \ggg 0$ is the largest possible Fermat prime. Then the Hausdorff dimension of $C(p)$ is $\log (2^N)/ \log (2^{N+1}-1) \sim N/N+1$ which tends to unity (smoothly) as $N \rightarrow \infty$. 

Now such $C(p)$ contain as many points as does the unit interval in which $C(p)$ is embedded (an uncountable infinity). Hence, for $y \in C(p)$, there are as many perturbations $\delta_{C(p)}: y \mapsto y'$ which leave $y' \in C(p)$ as perturbations $\delta_{\cancel{C(p)}}: y \mapsto y''$ for which $y'' \notin C(p)$. Let us describe the former as `geometrically constrained' perturbations, the latter as `geometrically unconstrained' perturbations. As discussed below, both play a key role in understanding why IS theory is not conspiratorial. 

By construction, IS theory is robust to geometrically constrained perturbations. We can describe this robustness using a tool which is commonplace in number theory - but less so in physics - p-adic analysis  \cite{Robert}. The p-adic number systems are relevant here because there exists a well-known homeomorphism, $F_p$, between the ring of $p$-adic integers and $C(p)$, defined by \cite{Vallin}
\begin{equation}
\label{adic}
F_p \left(  \sum_{k=0}^{\infty} a_k p^k \right )=  \sum_{k=0}^{\infty} \frac{2a_k}{(2p-1)^k+1}    
\end{equation}
For example, with $p=2$, (\ref{adic}) is a bijection between the dyadic integers and the Cantor Ternary Set. The p-adic integer $x$ is equipped with a p-adic norm $|x|_p$. This in turn defines a metric $|x_1-x_2|_p$ between two p-adic integers $x_1$ and $x_2$. This metric can be used to define the notion of distance $D(y_1,y_2)$ between two points on $C(p)$:
\begin{equation}
D(y_1,y_2)=|x_1-x_2|_p
\end{equation}
where $y_1=F_p(x_1)$, $y_2=F_p(x_2)$. In particular, since $|x_1-x_2|_p \le 1$ for any two p-adic integers, so also $D(y_1, y_2) \le 1$ for any two points on the Cantor set. Now, let $x$ be a $p$-adic integer and $y=F_p(x)$ the corresponding point on $C(p)$. If we perturb $x$, i.e. add to $x$ some $p$-adic integer $\delta_{I}$ such that $|\delta_{I}|_{p} \ll 1$, then not only $y'=F_p(x+\delta_I) \in C(p)$, but also the smaller is $\delta_I$, the closer $y'$ is to $y$ both in the Euclidean metric $E$ of the Euclidean space in which $C(p)$ is embedded, and in $D$. 

Now the p-adic integers $\mathbb{Z}_p$ are readily extended to the p-adic rationals $\mathbb{Q}_p$, into which the rationals can be embedded. The p-adic metric extends straightforwardly onto $\mathbb{Q}_p$ and the corresponding extension of $D$ encapsulates the essential physics of the Cosmological Invariant Set Postulate: that a geometrically unconstrained perturbation is a large-amplitude perturbation. To see this, let us add to $x$ a rational number $\delta_{Q}$, such that $\delta_Q$ belongs to $\mathbb{Q}_p$, but not $\mathbb{Z}_p$. Then $y''=F_p(x+\delta_Q)$ does not lie on $C(p)$.  The distance $D(y'', y)$ is, by definition, $|\delta_Q|_p$. Now the p-adic norm of a p-adic rational which is not a p-adic integer is necessarily greater than or equal to $p$. Since $p \ggg 0$, we have the result that $D(y_1, y_2)\ll 1 \Rightarrow E(y_1, y_2) \ll 1$, whilst $E(y_1, y_2) \ll 1 \nRightarrow D(y_1, y_2) \ll 1$.  

In summary, IS theory is robust to p-adic noise, and perturbations which are small amplitude in the $E$ metric are not necessarily small amplitude in the more physically relevant $D$ metric. We discuss the important physical implications of this in the next Section. 

Informally, the real numbers can be considered the singular limit of the $p$-adic numbers, at $p=\infty$ \cite{Neukirch}. This is relevant in showing that the complex Hilbert Space of quantum theory is the singular limit of IS theory at $p=\infty$ \cite{Palmer:2015b}. 

\section{Conspiracy and CHSH}
\label{conspire}

The discussion about p-adic numbers  is directly relevant to the issue of conspiracy. It is sufficient to discuss the example raised by Bell himself \cite{Bell} to be able to tackle the concept of conspiracy in IS theory. Imagine a CHSH experiment, where $\mathbf{a}$ and $\mathbf{b}$ take the discrete binary orientations $(\mathbf{a}_1, \mathbf{a}_2)$, $(\mathbf{b}_1, \mathbf{b}_2)$ respectively. Let us suppose that these values are set by two independent pseudo-random number generators, whose outputs are sensitive to the values $\alpha, \beta \in \{0,1\}$ of (say) the millionth bits of the input variables, respectively. Now in order to violate Measurement Independence,  the probability distribution $\rho(\lambda| \alpha, \beta)$ of the ontic variables must also depend on $\alpha$ and $\beta$. Using traditional thinking, any such dependence seems implausible; whilst $\alpha$ and $\beta$ determine $\mathbf{a}$ and $\mathbf{b}$ respectively, it seems hard to imagine - Bell at least found it hard to imagine - that they are the crucial pieces of information for any other distinct purpose, such as constraining the ontic variables. This is the basis of the idea that an implausible `conspiracy' (between $\alpha$, $\beta$ and $\lambda$) is needed to violate Measurement Independence. 

However, let us try to analyse this issue using IS theory. Suppose, $\alpha=0$ and consider the counterfactual question Q: What would have been the outcome of the measurement on a particle with ontic variable $\lambda_0$ if instead $\alpha=1$? If the counterfactual perturbation which takes $\alpha$ from 0 to 1, keeping $\lambda_0$ fixed, is an example of a geometrically unconstrained perturbation (c.f. Section \ref{ft}), then, with probability one, this perturbation will take the state of the universe $X_U$ to a perturbed state $X'_U$ off the invariant set. As such the distance between $X_U$ and $X'_U$ in the physically based metric $D$ is large, even if the perturbation appears inconsequentially small with respect to the Euclidean embedding space metric. In this situation the value of $\alpha$ certainly is a crucial piece of information, not only for determining $\mathbf{a}$, but for ensuring the existence of every atom in the universe! In this sense, the fact that $\alpha$ `only' defines the millionth digit of the input variable is a complete red herring; in the $D$ metric, perturbing the millionth digit keeping $\lambda_0$ fixed may correspond to a very large perturbation. 

Conversely, if the counterfactual perturbation above is a geometrically constrained perturbation, then it will map the state of the universe $X_U$ to a perturbed state $X'_U$ on the invariant set. In this situation, the smallness of the distance between $X_U$ and $X'_U$  in the Euclidean metric implies the smallness of distance in the $D$ metric. In this situation, the intuition that $\alpha$ cannot be the crucial piece of information for any other distinct purpose (than for setting $\mathbf{a}$) is correct. If $I_U$ were the whole of Euclidean state space, then all perturbations would be of the geometrically constrained type and Bell's intuition would have been correct. This is the case for traditional hidden-variable theories (e.g. Bohmian theory). It is not, however, for IS theory: no matter how large is $p$ (i.e. no matter how close the dimension of $C(p)$ is to an integer), the measure of $I_U$ is strictly zero in the Euclidean space in which it is embedded.

Is the perturbation corresponding to the counterfactual question Q above, geometrically constrained or geometrically unconstrained? The answer depends on an analysis using IS theory. This is given in reference \cite{Palmer:2015a} for the CHSH experiment. For  given $\lambda_0$, if the pair of orientations are in reality $(\mathbf{a}_i, \mathbf{b}_j)$, where $i, j \in \{1,2\}$, then the state of a counterfactual universe where the orientations are either $(\mathbf{a}_k, \mathbf{b}_j)$, $k \ne i$ or $(\mathbf{a}_i, \mathbf{b}_k)$,  $k \ne j $,  $k \in \{1,2\}$, lies off $I_U$, i.e. the corresponding perturbations are geometrically unconstrained. Now the CHSH inequality 
\begin{equation}
\label{chsh}
A=|\mathrm{Corr}_{\rho}(a_1, b_1) - \mathrm{Corr}_{\rho}(a_1, b_2)|+|\mathrm{Corr}_{\rho}(a_2, b_1)+\mathrm{Corr}_{\rho}(a_2, b_2)| \le 2
\end{equation}
involves all four pairs of possible settings for $(\mathbf{a}_i, \mathbf{b}_j)$. For given $\lambda_0$, if any one pair of settings is associated with a state of $U$ on $I_U$ then the other two settings are not associated with a state of $U$ on $I_U$. That is to say, $A$ is necessarily undefined in IS theory. Hence IS theory is not constrained by the CHSH inequality. 

As with the original Bell inequality, IS theory asserts that what is actually estimated when the CHSH inequalities are found to be violated experimentally is the inequality
\begin{equation}
\label{chsh2}
A'= |\mathrm{Corr}_{\rho_1}(a_1, b_1) - \mathrm{Corr}_{\rho_2}(a'_1, b_2)|+|\mathrm{Corr}_{\rho_3}(a_2, b'_1)+\mathrm{Corr}_{\rho_4}(a_2, b_2)|\le 2
\end{equation}
where $\mathrm{Corr}_{\rho}(a_i, b_j)=\cos \theta_{a_i b_j}$ and $a'_1=a_1$, $b'_1=b_1$ to within the necessarily finite precision of the measuring instruments, such that all of $\cos \theta_{a_1 b_1}$, $\cos \theta_{a'_1 b_2}$, $\cos \theta_{a_2 b'_1}$ and $\cos \theta_{a_2 b_2}$ are describable by $N$ bits. Importantly, from (\ref{chsh}) and (\ref{chsh2})
\begin{equation}
\label{a}
A \ne A':
\end{equation}
the left hand side of (\ref{a}) is undefined, whilst the right-hand side is not. Put another way, there exists no experimental protocol (on the invariant set) from which $A$ can be estimated. By contrast, $A'$ can be estimated experimentally: one performs four separate sub-experiments, one for each correlation. Since, by construction, each sub-experiment is an element of physical reality, it must be the case that each of $\cos \theta_{a_1 b_1}$, $\cos \theta_{a'_1 b_2}$, $\cos \theta_{a_2 b'_1}$ and $\cos \theta_{a_2 b_2}$ is describable by $N$ bits. For such measurements, $A'$ can exceed 2, even though it is not the case that $A>2$ in IS theory.

The argument above is reminiscent of the finite-precision nullification of the Kochen-Specker theorem  \cite{Meyer:1999}. In the present case, however, such nullification is based on an underlying physical premise: the Cosmological Invariant Set postulate.  

\section{Free Will}
\label{free}

A compatibilist definition of free will \cite{Kane} implies that we are free when there is an absence of constraints or impediments preventing us from doing what we want to do. Of course, in the sense that we are constrained by the laws of physics, we are never completely free. I may have the desire to fly like a bird by flapping my arms up and down, but the laws of physics will prevent my realising this desire. Hence, as a definition of experimenter free will, let us say that experimenters have free will when there is an absence of constraints or impediments preventing them from doing what they choose to do, providing these choices are consistent with the laws of physics (here the word `choice' is presumed to describe the result of some complicated set of neurological processes, triggered by input from the senses)

Let us suppose Alice and Bob choose orientations $\mathbf{a}$ and $\mathbf{b}$ with relative angle $\theta_{ab}$. After they have made this choice, we ask them to write down their choices in terms of the angular coordinates for $\mathbf{a}$ and $\mathbf{b}$. There are two possibilities: either the corresponding $\cos \theta_{ab} \in \mathbb{Q}_2(N)$, or the corresponding $\cos \theta_{ab} \notin \mathbb{Q}_2(N)$. If the former, then IS theory is able to satisfy their choices directly. However, if $\cos \theta_{ab} \notin \mathbb{Q}_2(N)$ then, providing $N$ is large enough, there will exist a $\delta \theta$ which is smaller than the finest possible angular resolution of Alice and Bob's measuring instruments such that $\cos \theta'_{ab} \in \mathbb{Q}_2(N)$ where $|\theta'_{ab} - \theta_{ab}| < \delta \theta$.  In IS theory, $N$ is assumed to be sufficiently large that this condition is satisfied. Hence, Alice and Bob's choices can be accommodated by IS theory providing we recognise that the laws of physics prevent an exact realisation of their choices of measurement orientation if they are overly precise. The crucial question is whether, as experimenters, Alice and Bob need be aware of this theoretical restriction as a practical restriction on the types of experiment they may wish to perform. In (the complex Hilbert Space of) quantum theory, measurement statistics vary continuously with $\theta_{ab}$. In IS theory these measurement statistics come as close as one likes to varying continuously, providing $N$ is large enough - this is because the fractal dimension of $I_U$ becomes as close as one likes to an integer value for large enough $N$. For large enough $N$, Alice and Bob cannot distinguish between the strictly continuous variation of measurement statistic predicted by quantum theory, and the almost continuous variation of measurement statistic predicted by IS theory. (It is crucially important to note, however, that because quantum theory is the singular limit and not the smooth limit of IS theory, we can make $N$ as large as we like without the state space of  IS theory ever approximating the algebraically closed state space of quantum theory: the counterfactual incompleteness of IS theory holds for all finite $N$, no matter how large).

In conclusion, IS theory allows experimenters unfettered freedom to choose measurement orientations as they wish. Experimenters have free will in every practical sense of the phrase!
 
\section{Retrocausality, Non-computability and Predictability}
\label{retro}

Let us temporarily move away from the Bell inequalities and consider a delayed choice experiment in a Mach-Zehnder interferometer. Consider a time $t_0$ when a photon has just passed through the first beam splitter of the interferometer, but when the experimenter has yet to decide whether to insert the second beam splitter. If the experimenter puts in the second beam splitter, it is only at the later time $t_1>t_0$. The well-known delayed choice `paradox' is to understand how the photon knows at $t_0$ whether to behave like a wave or like a particle.  Is it retrocausality? In IS theory, there is no need to invoke retrocausality; the structure of the invariant set at $t_0$ is deterministic and has to be consistent with wave-like or with particle-like properties for the photon, consistent with the actual measurement made at $t_1$. At first sight, this does not seem to explain the paradox at all: if a theorist can deduce mathematically the structure of the invariant set at $t_0$, she can predict what experiment will be conducted at $t_1$. That is to say, the theorist will be able to tell the experimenter whether he will put the second beam splitter into the interferometer or not. Such an experimenter can be expected to be more than happy to prove the theorist wrong and do the opposite! Of course the answer to this `predictability' paradox is (according to IS theory) that whilst the structure of the invariant set  is certainly well defined at $t_0$, no finite computation (by theorist or computer) can reveal this structure - the theorist will be unable to reveal any useful information to the experimenter about what experiment will actually be performed at the later time $t_1$. Fractal invariant sets are formally non-computable \cite{Blum}. 

In the case of an EPR experiment,  one can imagine the pair of particles being produced at $t_0$ and the experimenters deciding the measurement orientations at $t_1$. The invariant set at $t_0$ has to be consistent with these orientations. However, no theorist can probe the invariant set at $t_0$ to somehow predict which orientations will be chosen by the experimenters. 

The notion of non-computability is not commonplace in physics. It arises in the current context because the invariant set is defined using the methods of global analysis (see \cite{Palmer:2015a}). One can draw on an analogy in GR of a concept which also makes explicit use of methods of global analysis (in space-time rather than state space).  The black-hole event horizon $\mathscr{H}^+$ is defined as the null boundary of light rays which escape to future null infinity. Consider a massive object $M$ orbiting a black hole well outside the event horizon. Suppose at $t=t_1$ (where $t$ labels a family of space-like hypersurfaces), an experimenter tosses a coin. If the coin lands heads, the experimenter propels $M$ into the black hole. If the coin lands tails, $M$ remains in orbit. Because $\mathscr{H}^+$ is defined by a global space-time condition, the position of $\mathscr{H}^+$ at $t_0<t_1$ also depends on the coin toss at $t_1$. For example, a null ray which appeared to be diverging from the black hole between $t_0$ and $t_1$ would become trapped at some time $t>t_1$ when $M$ had fallen into the black hole. This null ray, at $t_0$, would therefore belong to the event horizon. If one imagined that the position of the event horizon at $t_0$ could be calculated from the local curvature of space-time, then the correlation between the position of $\mathscr{H}^+$ at $t_0$ and the outcome of the coin toss could only be explained by assuming some form of backward causality or some implausible conspiracy. However, the position of $\mathscr{H}^+$ cannot be calculated in this way. Indeed the precise position of $\mathscr{H}^+$ cannot be calculated from any finite algorithm: it is technically non-computable. Of course, there is no paradox here once one realises that $\mathscr{H}^+$ is a globally defined (yet manifestly causal) concept. 
 
\section{Cosmology, Measurement and Quantum Gravity} 
\label{cosmo}

In IS theory, the Invariant Set Postulate is a top-down \cite{Ellis} constraint from cosmology to quantum physics. A key question is whether this posulate is consistent with our understanding of the structure of cosmological space-time, since the existence of a measure-zero fractal invariant set indicates a departure from Hamiltonian dynamics (and, by the Liouville theorem, state-space volumes need not be conserved if the underlying dynamics is non-Hamiltonian). 

Crucially, departures from Hamiltonian structure need only occur in very localised areas of state space, in order that the invariant set is globally fractal.  We speculate here that the required departures from non-Hamiltonian structure are specifically associated with localised regions of state space associated with (classical) space-time singularities. In particular, Penrose \cite{Penrose:2010} argues for a reduction of state-space volume associated with dynamical evolution through a space-time singularity (based on the notion of information loss in black holes). Such contraction would correspond to a convergence of state-space trajectories of the type seen generically in those nonlinear dynamical systems which exhibit fractal invariant sets. If these ideas are correct, then whilst it may well be the case that quantum physics is needed to understand the nature of space-time singularities, it is even more true that space-time singularities are needed to understand the nature of quantum physics!

Overall, IS theory requires a quasi-cyclic cosmology evolving on a measure-zero invariant set. The neighbouring space-time trajectories on $I_U$ (whose statistical properties are described by IS theory) should therefore be thought of, not as `other worlds', but rather as our unique world at earlier or later epochs. The contraction of state space at each final space-time singularity would be enough to reset cosmological entropy ahead of the big bang for the next epoch. It is conceivable that dark energy is required to ensure that the invariant set $I_U$ is not a trivial fixed point or simple limit cycle (both of which would be too super-deterministic to explain quantum physics without conspiracy). 

Because IS is a geometric theory strongly linked to a dynamical systems approach, and because GR can be written as a dynamical systems theory (e.g. the ADM formulation), the synthesis of GR with IS theory can be expected to be much less problematic that with quantum theory. Indeed the basis of measurement in IS theory is the `clumping' of state-space trajectories on $I_U$ into discrete classes. This allows trajectories to be given a symbolic labelling, and the statistics of such labelling is the basis of the injection into the complex Hilbert Space, as mentioned above \cite{Palmer:2015b}. In IS theory, this state-space clumping can be considered a manifestation of the phenomenon we refer to as `gravity'. This already leads to certain predictions. For example, an emergent property of IS theory is that vacuum energy will not couple directly to the gravitational field \cite{Palmer:2014} \cite{Palmer:2015b}. More radically, IS theory predicts there is no such thing as a graviton. As mentioned above, new perspectives on dark energy are also emergent. 

The very notion that quantum physics should be emergent from a cosmological - and hence overtly gravitational -  constraint, suggests it may be wrong to search for a `quantum theory of gravity'; we should instead be seeking a `gravitational theory of the quantum'.
 
\section{Conclusions}
\label{conc}

Invariant Set theory is based on the premise that the Universe as a whole can be considered a dynamical system evolving precisely on a fractal invariant set $I_U$ in its state space. In this paper we have not discussed the precise nature of this set - it must, for example,  incorporate quaternionic structure to be consistent with the quantum physics of spin. Rather we have focussed on the consequences of an emergent constraint (call it $C$) arising from this invariant set structure - specifically that the cosine of relative measurement orientations must be describable by finite $N$ bits, where $N \gg 0$ is a parameter describing the fractal dimension of $I_U$. 

The key result exploited in this paper is that for an arbitrary triangle on the sphere, where two sides satisfies $C$, and one of the interior angles is a rational multiple of $\pi$, then the third side cannot satisfy $C$. We have applied $C$ to the Bell inequality and shown that it allows Measurement Independence to be partially violated - without violating local causality or realism. This triangle property also applies to the interpretation of other key quantum phenomena such as such as quantum interferometry \cite{Palmer:2015b} and the sequential Stern-Gerlach experiment \cite{Palmer:2014}. In the latter case, for example, the triangle property provides an explanation of what otherwise is `explained' by the Heisenberg Uncertainty Principle - that simultaneous incompatible measurements of spin are impossible. As a result of the discussion it is plausibly the case that IS theory can provide a fully realistic and locally causal description of all quantum physics. 

The triangle property above arises from an elementary application of number theory. Number theoretic concepts also arise in showing why IS theory is robust to noise. In particular, we have utilised the homeomorphism between the ring of $p$-adic numbers and the Cantor Set $C(p)$ which describes the structure of $I_U$ in directions transverse to the state-space trajectories, where $p=2^N+1$. IS theory is robust to $p$-adic noise. In general, the theory of p-adic numbers is not familiar to physicists, though in number theory the p-adics are used as frequently as the more familiar reals. Indeed, it was noted above that quantum theory itself can be considered the singular limit of IS theory when $p= \infty$ (consistent with the reals being considered the singular limit of the p-adics when $p=\infty$). It is worth reflecting on other singular limits in physics. For example, as mentioned, the inviscid Euler equations are the singular limit of the viscous Navier-Stokes equations for infinite Reynolds number. For many purposes, the Euler equations provide a good approximation to the Navier-Stokes equations for high Reynolds number flow. However, on occasion the Euler equations prove disastrously unrealistic. They predict, for example, that aeroplanes could never fly!

Similarly, there may be circumstances where quantum theory proves disastrously unrealistic, such as in the cosmological domain where quantum and gravitational physics are both important. The grossly unrealistic estimates of the cosmological constant based on quantum field theoretic estimates of vacuum energy may be an example (in IS theory, such small fluctuations are not gravitationally coupled \cite{Palmer:2014}). IS theory suggests that it may be misguided to seek a quantum theory of gravity. Rather, the constraint played by the overtly gravitational cosmological invariant set in IS theory suggests that the unification of gravitational and quantum physics will instead be found in a gravitational theory of the quantum. 

\section*{Acknowledgements}
My thanks to Shane Mansfield for helpful discussions on quantum foundational issues, to Kristian Strommen for help about p-adic numbers, and to the anonymous referees of this paper for many insightful comments. 
\nocite{*}
\bibliographystyle{eptcs}
\bibliography{mybibliography}
\end{document}